\documentclass[twocolumn]{aastex631}
\usepackage[utf8]{inputenc}
\usepackage[T1]{fontenc}
\usepackage{graphicx} 
\usepackage{amsmath}
\usepackage[sort&compress]{natbib}
\usepackage{float}
\usepackage{xspace}
\usepackage{appendix}

\newcommand{\degs}[1]{#1^{\circ}}

\newcommand{\sv}{\langle \sigma v \rangle}

\usepackage{hyperref}
\usepackage{tablefootnote}

\begin{document}
\title{Legacy analysis of Milky Way dwarf spheroidal satellite galaxies: an update}

\author[0009-0009-8180-2323]{Antonio Circiello}
\affiliation{Department of Physics and Astronomy, 
Clemson University,
Clemson, SC 29631, USA}

\author[0000-0003-2759-5625]{Mattia Di Mauro}
\affiliation{Istituto Nazionale di Fisica Nucleare, Sezione di Torino, 
Via P. Giuria 1, 10125 Torino, Italy}

\author[0000-0002-6584-1703]{Marco Ajello}
\affiliation{Department of Physics and Astronomy, 
Clemson University, 
Clemson, SC 29631, USA}

\author[0000-0002-6774-3111]{Christopher Karwin}
\affiliation{Department of Physics and Astronomy, 
Clemson University, 
Clemson, SC 29631, USA}

\author[0000-0001-8251-933X]{Alex Drlica-Wagner}
\affiliation{Fermi National Accelerator Laboratory, P.O.\ Box 500, Batavia, IL 60510, USA}
\affiliation{Kavli Institute for Cosmological Physics, University of Chicago, Chicago, IL 60637, USA}
\affiliation{Department of Astronomy and Astrophysics, University of Chicago, Chicago, IL 60637, USA}

\author[0000-0002-3849-9164]{Miguel Á. Sánchez-Conde}
\affiliation{Instituto de F\'isica Te\'orica UAM-CSIC, Universidad Aut\'onoma de Madrid, C/ Nicol\'as Cabrera, 13-15, 28049 Madrid, Spain}
\affiliation{Departamento de F\'isica Te\'orica, M-15, Universidad Aut\'onoma de Madrid, E-28049 Madrid, Spain}

\begin{abstract}
    Studies of \textit{Fermi}-Large Area Telescope (LAT) data coincident with dwarf spheroidal satellite galaxies (dSphs) of the Milky Way (MW) have put the most stringent constraints on models of annihilating dark matter (DM) with candidate masses in the GeV-TeV range.
    Recent results found the presence of small, local significance excesses from these targets, at the 2-3$\sigma$ level.
    However, these excesses disagree on the predicted properties of the DM candidate, and their significance vanishes when considering the correction factors for the number of trials.
    In this work, we apply key improvements to the analysis of the dSphs.
    We use stricter cuts on the data, implement a method to adaptively model the background, and assume an updated framework for DM annihilation.
    We find that our improved background modeling leads to a better agreement between the model and the data. This produces an increase in the local and global significance of the dSphs excess compared to previous studies.
    Finally, we find that the DM properties obtained in this work are less dependent on the sample of dSphs being considered compared to previous studies, while remaining in agreement with the predictions from the Galactic center excess observed by the \textit{Fermi}-LAT and the antiproton excess observed by the Alpha Magnetic Spectrometer (AMS-02). Considering our improvements, a future significant increase in the number of dwarfs may lead to a definitive confirmation or exclusion of the DM interpretation of the Galactic center excess.
    \end{abstract}

\section{Introduction}
\label{sec:intro}

\noindent Strong cosmological and astrophysical evidence indicates that dark matter (DM) makes up roughly $25\%$ of the energy density of the Universe \citep{Planck+18}, yet its nature remains elusive.
Among the multitude of viable candidates, weakly interacting massive particles (WIMPs) are a well-motivated class of particles that naturally arise as an extension of the Standard Model \citep{Bertone+05, Bergstrom+2012, Arcadi+2018}.
In the standard thermal relic scenario, the present-day abundance of WIMPs is determined by their annihilation cross section and mass at the time of freeze-out.
For cross sections characteristic of the weak interaction and masses in the GeV-TeV range, the predicted relic abundance would roughly match the observed abundance of DM in the Universe \citep{Jungman+96, Bergstrom+00}.
Furthermore, due to its large mass, the annihilation of WIMPs into Standard Model particles is expected to produce secondary photons in the $\gamma$-ray band \citep{Bertone+05, Cirelli+11, Arina+24}.
The Large Area Telescope (LAT) on board the \textit{Fermi} satellite \citep{fermiLAT}, due to its sensitivity in the 100 MeV to >1 TeV range, has been leading the search for these $\gamma$-ray signatures of DM annihilations, under the WIMP hypothesis.
Although no unambiguous detection has been achieved, this continued observational effort has placed strong constraints on the WIMP annihilation cross section over a broad mass range, limiting the viability of the thermal relic hypothesis, especially at low WIMP masses \citep[e.g., ][]{Ackermann+11, Geringer-Sameth+11, Ackermann+15, Geringer-Sameth+15b, Albert+17, DiMauro+21, Thorpe+21, DiMauro+23, AMD+24}.
Gravitational signatures of DM show that it is almost ubiquitous throughout the Universe.
However, the viability of a target for searches of $\gamma$-ray signatures of DM annihilation depends on various factors, like DM density, astrophysical backgrounds, and modeling uncertainties.
A wide range of studies have been performed on \textit{Fermi}-LAT data coincident with varied classes of targets, like the Milky Way (MW) Galactic center \citep{Hooper+11, Calore+15, Daylan+16, Karwin+17, DiMauro:2021raz, DiMauro+21}, the Andromeda galaxy \citep[M31, ][]{Ackermann+17, Karwin+19, Karwin+21}, clusters of galaxies \citep{Colafrancesco+06, Lisanti+18, Thorpe+21, DiMauro+23}, the isotropic $\gamma$-ray background \citep[IGRB, ][]{Bergstrom+01, Bringmann+14, Ajello+15, DiMauro:2015tfa, Fermi+15_IGRB} and dwarf spheroidal satellite galaxies \citep[dSphs, ][]{Abdo+10, Ackermann+11, Ackermann+15, Drlica-Wagner+15, Geringer-Sameth+15b, Albert+17, Calore+18, DiMauro:2020uos, Hoof+20, DiMauro:2021raz, AMD+24}.
The Galactic center provides, in principle, the strongest expected flux from DM annihilation, due to its proximity and high expected density of DM. 
On the other hand, the signal in this area of the sky is dominated by the bright emission of our own Galaxy, rendering searches for DM very complex.
However, careful studies of the \textit{Fermi}-LAT data coincident with the Galactic center have highlighted the presence of a seemingly diffuse excess $\gamma$-ray emission at the GeV energies \citep{Hooper+09}, known as the Galactic center excess (GCE), which is not conclusively associated with any known population of sources residing in the region. 
The GCE has been extensively studied; however, it is still a matter of debate whether it represents the first detection of a truly diffuse signal from DM annihilation \citep{Hooper+11, Calore+15, Daylan+16, Fermi+17, Karwin+17, DiMauro:2021raz, DiMauro+21}, or it is rather produced by a large population of unresolved point sources, like millisecond pulsars \citep{Abazajian+11, Bartels+16, Lee+16, Malyshev+25, Kalambay+26}.
M31 shares similar advantages and challenges to the Galactic center, being hosted in a nearby (although much farther than the Galactic center) massive halo with a relatively well-known morphology, but also being polluted by bright $\gamma$-ray foregrounds \citep{Ackermann+17, Karwin+19, Karwin+21}.
Galaxy clusters would have large total masses of DM, but yield relatively low signals due to the large distances \citep{Colafrancesco+06, Lisanti+18, Thorpe+21, DiMauro+23}.
Studies of the IGRB are based on the principle that the cumulative emission from DM annihilation in the Universe could make up for a measurable portion of the diffuse background.
Comparing the diffuse emissions from known populations of  $\gamma$-ray sources to the IGRB observed by the LAT could help characterize this diffuse DM emission. 
However, these studies have to deal with high uncertainties in the luminosity functions of the known populations of $\gamma$-ray sources and their propagation in the process of combining the different diffuse emissions together, limiting the precision that can be achieved \citep{Bergstrom+01, Bringmann+14, Ajello+15, DiMauro:2015tfa}.

Finally, the dSphs, focus of this study, are among the most appealing targets due to their proximity and low astrophysical backgrounds.
These nearby compact galaxies are characterized by a large mass-to-light ratio and minimal intrinsic $\gamma$-ray emission from other processes.
For this reason, when they lie in regions of the sky with low $\gamma$-ray foregrounds or backgrounds, they provide a particularly clean probe of DM.
Due to these advantages, the population of dSphs has been thoroughly investigated, leading to some of the most robust and stringent constraints on the properties of WIMPs \citep{Abdo+10, Ackermann+11, Geringer-Sameth+11,Ackermann+15, Drlica-Wagner+15, Geringer-Sameth+15b, Albert+17, Calore+18, DiMauro:2020uos, Hoof+20, DiMauro+21}.
The most recent analysis of LAT data coincident with the dSphs population by \citet{AMD+24} highlights the presence of a small excess, although only locally significant (at the $2 -3 \; \sigma$ level). 

Even in relatively clean targets such as the dSphs, mismodeling of the background can significantly impact the results, due to the faint nature of the expected emission. 
In this work, we present an update to the latest study by \citet{AMD+24}, implementing adjustments to the analysis to improve the modeling of the background.
Additionally, we implement an updated framework for DM annihilation.
\cite{AMD+24}, as many previous searches of signatures of DM annihilation in the LAT data, uses the PPPC4DMID package presented in \cite{Cirelli+11}.
An updated package, CosmiXs, that includes a more careful treatment of the electroweak interactions involved in the annihilation processes, has been recently developed by \cite{Arina+24}, and is used for the first time to analyze the dSphs in this study.

The paper is organized as follows:
\S \ref{sec:annihilation} outlines the properties of the $\gamma$-ray emission from DM annihilation; \S \ref{sec:sample} summarizes the criteria used for the selection of the dSphs sample; \S \ref{sec:analysis} describes the analysis of \textit{Fermi} data coincident with the dSphs; \S \ref{sec:updates} describes in detail the differences from the old versions of the analysis and their impact on the results; \S \ref{sec:results} presents the results of the analysis; \S \ref{sec:conclusion} presents our final conclusions on the results.

\section{Dark Matter annihilation}
\label{sec:annihilation}
\noindent The expected $\gamma$-ray flux from DM annihilation in a dSph can be written as:
\begin{equation}
    \label{eq:flux}
    \frac{d\Phi_\chi}{dE} = J \times \frac{1}{4\pi}\frac{\sv}{2 M_{\chi}^2} \sum_i \beta_i \frac{dN_i}{dE},
\end{equation}
where $\sv$ is the thermally averaged cross-section for annihilation, and $M_\chi$ is the mass of the candidate DM particle.
The sum is calculated over all possible channels of annihilation $i$, with branching ratio $\beta_i$ and differential photon yield per annihilation event $dN_i/dE$.
This factor contains all the information on the assumed model of annihilating DM.
The $dN_i/dE$ in the different channels of annihilation assumed in this work are taken from the "CosmiXs" annihilation package presented in \cite{Arina+24}.
We perform the analysis considering the bottom quark ($b\bar{b}$) and $\tau$ lepton ($\tau^{+}\tau^{-}$) annihilation channels separately with a branching ratio of $\beta_i = 1$, as being representative of hadronic and leptonic annihilation channels, respectively.
The remaining term in Eq. \ref{eq:flux} is the $J$-factor, $J$, defined as \citep{Bergstrom+98}:
\begin{equation}
    \label{eq:j-factor}
    J = \int_{\Delta\Omega}\int_{l.o.s.} \rho^2_\chi(l, \Omega)  d\Omega d l,
\end{equation}
which is the square of the matter density of DM in the system ($\rho_\chi$) integrated over the line of sight (l.o.s.) and the underlying solid angle ($\Omega$).
The $J$-factor of a dSph is independent of the particular DM particle model assumed and can be estimated from the kinematic properties of the system being analyzed.
Direct measurements of the $J$-factor are available for a good portion of the observed dSphs.
These values are obtained by studying the dynamics of the member stars of the dSphs to trace the underlying gravitational potential of the host halo \citep{Geringer-Sameth+15b}.
For a significant fraction of dSphs, however, measurements of the $J$-factor are unavailable due to either technological or physical constraints (e.g., limited availability of observations, limited telescope resolution or limited number of member stars).
For these systems, lacking precise spectroscopic measurements, the $J$-factor can still be estimated using either limited kinematic information or their photometric properties \citep{Drlica-Wagner+15, Evans+16, Pace+19}, assuming that they are DM-dominated.
In particular, \cite{Pace+19} presented the most updated iteration of the empirically derived correlation between a dSphs $J$-factor and its kinematic properties as:

\begin{align}
    \label{eq:kine-J}
    \frac{J(0.5^\circ)}{\mathrm{GeV^2 cm^{-5}}} =&\; 10^{17.87} 
    \left(\frac{\sigma_{\mathrm{l.o.s.}}}{5 \;\mathrm{km\,s^{-1}}}\right)^{4} \nonumber \\
    &\times \left(\frac{d}{100 \;\mathrm{kpc}}\right)^{-2}
    \left(\frac{r_{1/2}}{100 \; \mathrm{pc}}\right)^{-0.5}
\end{align}

where $\sigma_\mathrm{l.o.s.}$ is the velocity dispersion of the member stars along the line of sight in $\mathrm{km}\,\mathrm{s}^{-1}$, $d$ is the heliocentric distance from the target in kpc, and $r_{1/2}$ is the azimuthally averaged physical half-light radius of the dSphs.
In the absence of even this limited kinematic information, reflected in the value of $\sigma_{l.o.s.}$, the $J$-factor can be estimated from the photometric properties of the dSphs, as:
\begin{align}
    \label{eq:photo-J}
    \frac{J(0.5^\circ)}{\mathrm{GeV^2 cm^{-5}}} =&\; 10^{18.17} 
    \left(\frac{L_V}{10^4 L_\odot}\right)^{0.23} \nonumber \\
    &\times \left(\frac{d}{100 \;\mathrm{kpc}}\right)^{-2}
    \left(\frac{r_{1/2}}{100 \; \mathrm{pc}}\right)^{-0.5}
\end{align}
where $L_V$ is the visual luminosity of the target.
We use these scaling relations for the dSphs that lack a direct measurement of the $J$-factor.
For these cases, we assume an uncertainty ($\sigma_J$) on the $J$-factor of 0.6 dex, representative of the expected measurement uncertainty.
While this is a standard choice for dSphs analyses, \cite{Albert+17} details the effects of different assumptions for $\sigma_J$ on the results, showing how they impact the uncertainty on the constraints for the annihilation cross-section.
Additionally, we note that the choice of the $J$-factor estimates depends on the underlying assumptions for the dSphs host haloes. 
The approach adopted here relies on empirical relations extrapolated from the properties of dSphs with measured $J$-factors. Alternatively, \cite{Ando+20} and \cite{Horigome+23} have shown how incorporating information from models of satellite galaxy formation can lead to reduced $J$-factor estimates for the faintest dSphs, and in turn to less stringent upper limits for $\sv$ by a factor of 2 to 7.

Finally, we assume that the dSphs can be modeled as point-like sources. 
This assumption is motivated by the estimated mass of the host haloes of these systems, and their expected projected size compared to the point-spread function of the \textit{Fermi}-LAT.
However, it has been noted that the spatial extension of these sources is likely to lower the sensitivity to DM annihilation signatures, by a factor that can reach up to 30$\%$ in the most extreme cases \citep[see ][for more in-depth discussions on this effect]{Drlica-Wagner+14, Ackermann+15, Geringer-Sameth+15b, DiMauro:2020uos}.

\section{Sample}
\label{sec:sample}
\noindent 
The list of detected MW satellites has been rapidly growing, thanks to the improvement in telescope technologies and the continued observational efforts of the community.
Discerning the DM-dominated dSphs from the DM-devoid globular clusters, however, can be challenging.
The primary way to do so is by measuring the physical properties of these systems and estimating the ratio of the stellar mass to the dynamical mass. 
Only for the brightest dSphs a DM density profile can be traced by measuring the mass profile at multiple radii.
For these reasons, samples for dSphs population studies often contain a mix of bona fide and putative dSphs.
For this study, we adopt the selection of 50 dSphs presented in \citet{AMD+24}.
Here we provide a brief summary of how the sample is selected, but we refer the reader to that study for more detailed information.

Of the $\sim 65$ detected dSphs of the MW, the ones that fall within the $95\%$ confidence radius of a source in the 4FGL-DR4 \citep{4fgl_dr3, Ballet+23} are excluded.
Similarly, dSphs with angular separation $<0.1^\circ$ from sources in the Roma-BZCat \citep{BZCat}, WIBRaLS \citep{WIBRaLS} and CRATES \citep{CRATES} catalogs are excluded. 
These cuts ensure that the remaining dSphs are devoid of significant contamination from known sources with detected or suspected $\gamma$-ray activity.

The remaining 50 dSphs are organized in three samples.
The `Measured' sample contains all the dSphs that have a spectroscopically measured $J$-factor.
The `Benchmark' sample contains all the dSphs from the Measured sample plus those whose $J$-factor is determined from their kinematic or photometric properties through equations \ref{eq:kine-J} or \ref{eq:photo-J}.
Finally, the `Inclusive' includes the two samples above plus eight special cases, which have either low Galactic latitude (i.e., $|b| < 15^\circ$) or are close to known radio sources \citep[see][for a detailed explanation of each special case]{AMD+24}.

In addition, following the same criteria used to select the samples of dSphs, we select $6500$ random high-Galactic-latitude (i.e., $|b| > 15^\circ$, to reduce Galactic foreground contamination) directions in the sky as our `blank fields'.
We select the blank fields so that any pair of directions has an angular separation of at least $1^\circ$.\footnote{This is to ensure that the regions of the sky selected for the \textit{Fermi} data analysis of the blank fields are fairly distinct.}
The analysis of the blank fields is used to calibrate the null hypothesis and quantify the significance of the results, as shown in sections \S~\ref{sec:analysis} and \S~\ref{sec:results}.

\section{Data Analysis and DM Likelihood}
\label{sec:analysis}
In this section, we outline the selections on the \textit{Fermi}-LAT data coincident with the dSphs and the details of the analysis.
For a thorough explanation of the changes from previous studies and their motivation, see \S \ref{sec:updates}.
For the analysis, we use the \texttt{fermipy} (v1.4.0) package and the underlying \texttt{fermitools} (v2.2.0) \citep{fermipy}.

We select $15.9$ years of \textit{Fermi}-LAT data taken between Aug 04, 2008 and Jun 21, 2024 in the 2 GeV to 1 TeV range.
Photons are selected from the P8R3 SOURCEVETO data \cite{Atwood+13, Bruel+18}, and we use the corresponding P8R3\_SOURCEVETO\_V3 instrument response functions.
We exclude photons with a zenith angle above $105^\circ$ to avoid Earth's limb contamination.
We model a region of interest (ROI) of $\degs{10} \times \degs{10}$ centred around each target.
The model includes the diffuse Galactic emission (\texttt{gll\_iem\_v07}) and the isotropic diffuse emission (\texttt{iso\_P8R3\_SOURCEVETO\_V3\_v1}), as well as point sources in the 4th data release of the 4th Fermi $\gamma$-ray light catalog (4FGL-DR4, \texttt{gll\_psc\_v35}) within $15^\circ$ from the target, and extended sources from the 14-year LAT templates.
We bin the ROI in eight logarithmically spaced bins per energy decade, and in spatial bins of $0.08^\circ$.
We take into account the energy dispersion for all components except the isotropic diffuse emission.

The model is fitted a first time leaving the isotropic and Galactic diffuse components free to vary, as well as the normalization and index of all the 4FGL sources in the ROI and of the target.
After this first fit, we use the \texttt{fermipy} \texttt{find\_sources()} method to account for the presence of additional sources that are not listed in the 4FGL-DR4, but might still be detected in the ROI, due to the additional years of data since the release of the catalog.
After a second fit, we compute the PS Map \citep{Bruel+21} and test statistics (TS) Map for the ROI, to evaluate the presence of residual significant model-to-data mismatches and modify the model of the ROI accordingly, as outlined in \S~\ref{sec:updates}.

We then compute the spectral energy distribution (SED) for each dSph. 
In each energy bin, the spectral index of the target is fixed at 2.0, while its normalization and the background emissions are left free to vary.
Finally, we convert the likelihood generated by the \texttt{fermipy sed()} function to the $M_\chi$ vs $\sv$ phase space, summing over the energy bins, and convolve it to a likelihood prior dependent on the $J$-factor value and uncertainty of the target.
The likelihood for a target with $J$-factor $J$, at given $M_\chi$ and $\sv$ is given by:
\begin{equation}
    \mathcal{L} (\sv, M_\chi, J) = \mathcal{L}_\Phi (\sv, M_\chi) \times \mathcal{L}_J(J,\sigma_J)
\end{equation}
where $\mathcal{L}_\Phi$ is the likelihood obtained from the comparison between the observed flux and its theoretical value from Eq. \ref{eq:flux} at given $M_\chi$ and $\sv$, combined over the energy bins ($E_i$):
\begin{equation}
    \mathcal{L}_\Phi = \prod_{E_i} \mathcal{L}_{\Phi_i} \left[ \frac{d\Phi_\chi}{dE}
    \left(M_\chi,\sv,E_i\right); E_i\right]
\end{equation}
with $\mathcal{L}_{\phi_i}$ likelihood in the $i$-th energy bin.
$\mathcal{L}_J$ is the prior dependent on the value of the $J$-factor and its uncertainty, defined as:
\begin{align}
    \mathcal{L}_J(J) =& \frac{1}{\mathrm{ln}(10) \sqrt{2\pi\sigma_J}J_{obs}} \nonumber \\
    &\times \exp{\left[- \left( 
    \frac{\mathrm{log_{10}}(J) - \mathrm{log_{10}}(J_{obs})}{\sqrt{2}\sigma_J}
    \right)^2\right]}
\end{align}
where $J_{obs}$ is the expected $J$-factor of the target, determined either through observation or the scaling relations above (Eq. \ref{eq:kine-J} and \ref{eq:photo-J}).
The likelihood is converted to a TS value as:
\begin{equation}
    \mathrm{TS} = 2\,\cdot\mathrm{ln}\left(\frac{\mathcal{L}_\Phi(\sv,M_\chi) \times \mathcal{L}_J}{\mathcal{L}_0}\right)
\end{equation}
where $\mathcal{L}_0$ is the null hypothesis likelihood, i.e., the likelihood when the target is removed from the model.
By scanning over a range of $M_\chi$ and $\sv$, we obtain a TS profile that peaks at the most likely values of these parameters, with higher values of the TS indicating increasing preference for the signal hypothesis.
All the TS profiles obtained in this way from the dSphs are then summed together to obtain joint constraints on $M_\chi$ and $\sv$.

However, $\mathcal{L}_0$ does not account for mismodeling of the background or the presence of hidden sources, and is therefore an inadequate characterization of the null hypothesis in this study, where the expected signal-to-noise ratio is low, and false positives are likely.
For this reason, we repeat the same analysis on random selections of the blank fields to better characterize the null hypothesis and quantify the significance of the results.
For each sample of dSphs, we consider 100000 random combinations without replacement of blank fields, selected to have the same number of targets as the sample being considered, and perform the combined likelihood analysis outlined above on each selection.
This allows us to characterize the TS fluctuations in the absence of a target, and use these distributions to evaluate the significance of the peak TS in the combined dSphs profiles.

The SED files and the TS profiles in the $M_\chi$ vs $\sv$ space for both the dSphs and the blank fields are available for download\footnote{\url{https://doi.org/10.6084/m9.figshare.32304966}}.

\section{Analysis Updates}
\label{sec:updates}
In this section, we discuss the differences between this analysis and the one presented in \cite{AMD+24} in more detail, the motivation behind them, and their impact on the results.

This work includes $\sim 2$ additional years of \textit{Fermi}-LAT data and updates the lists of sources included in the models to reflect the DR4 \citep[instead of DR3,][]{4fgl_dr3} of the 4FGL catalog.
As expected, this change has a marginal effect on the results, since the sensitivity to $\gamma$-ray signatures of DM annihilation achieved through the combined analysis of the dSphs, for a fixed size of the sample, scales as a combination of $\sim \sqrt{t}$, for background-limited searches, and $\sim t$, for signal-limited searches, as outlined in \cite{Charles+16}.

For the photon classification, we adopted the SOURCEVETO class, instead of SOURCE.
These two classes have the same behaviour at energies below 50 GeV.
However, at higher energies, the SOURCEVETO class has a lower background rate\footnote{For more details, see \url{https://fermi.gsfc.nasa.gov/ssc/data/analysis/documentation/Cicerone/Cicerone_Data/LAT_DP.html}}, 
slightly improving the fit quality above $E \sim 50$ GeV.
We can further improve the quality of the models by using TS Maps and, for the first time in a study of the dSphs, PS Maps to guide our choices in the data selection and modeling steps of the analysis.
The PS Map is a diagnostic tool for the goodness-of-fit of the \textit{Fermi}-LAT data that quantifies the mismatches between the integrated spectrum predicted by the model and the data in each pixel, with their significance.
The PS Map is able to detect both over- and under-predictions of the model compared to the data, with more reliable p-values than the ones obtained from a simple residual map.
The TS Map is a diagnostic tool used to evaluate the significance of a trial source of given morphology (in our case, a point-like source) added in each pixel of the ROI. The TS Map can be useful to identify and localize unmodeled $\gamma$-ray emission in the ROI. 
Compared to the PS Map, however, the TS Map is only sensitive to positive residuals.\\
When evaluating the PS Map of the ROIs modeled following the prescriptions in \cite{AMD+24}, we observed the presence of extended statistically significant mismatches between the model and the data. 
Such discrepancies are concentrated at low energies, where the background is less uniform and harder to model.
We therefore raise the minimum photon energy to $E > 2$ GeV, and perform the analysis as outlined in \S~\ref{sec:analysis}.
This choice for the energy threshold has minimal impact on the sensitivity to DM signals (see \S~\ref{apdx:sensitivity}).
When evaluating the PS Map with the new energy cut, the residuals are more uniform and lack any extended discrepancies.
However, after performing the analysis, we observed the presence of a few point-like excesses in the PS Map, whose significance can be evaluated using the TS Map.
We add point-like sources to the model in correspondence with these excesses, whenever the TS Map shows that they are detected at a TS $ > 16$ or above and have a minimum separation from the target of $0.5\deg$.
We then perform a last fit, leaving free to vary the parameters of the Galactic and isotropic diffuse, of the target, of the point-like excesses added from the PS Map and TS Map, and of the sources that are near local minima of the PS Map, where the model exceeds the data at a significance of at least $4\sigma$.

\begin{figure*}[t]
    \centering
    \includegraphics[width = 0.49\textwidth]{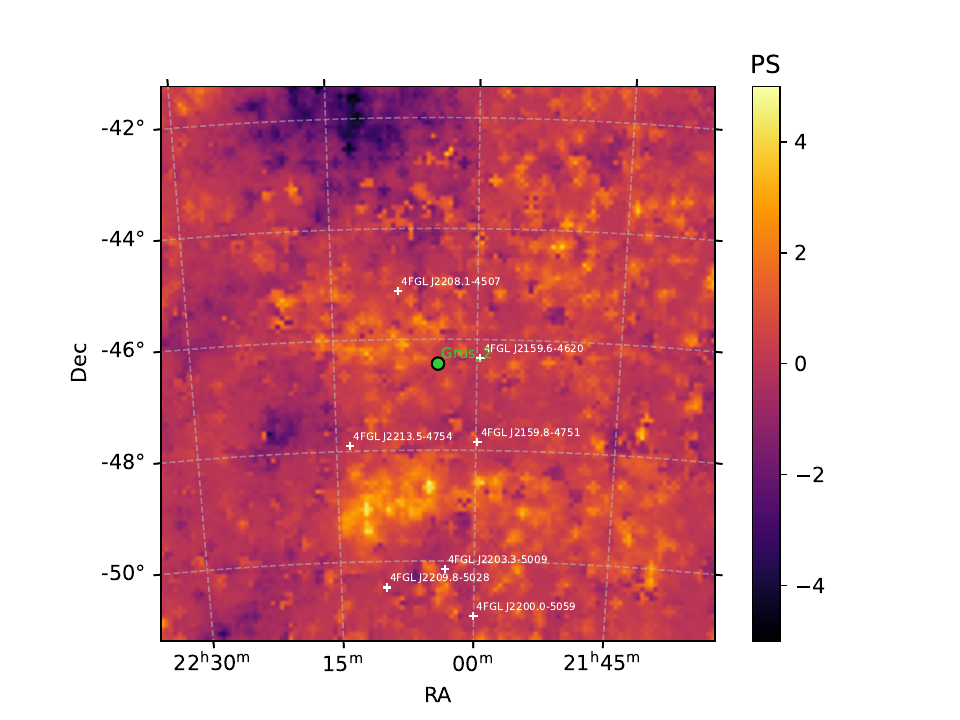}
    \includegraphics[width = 0.49\textwidth]{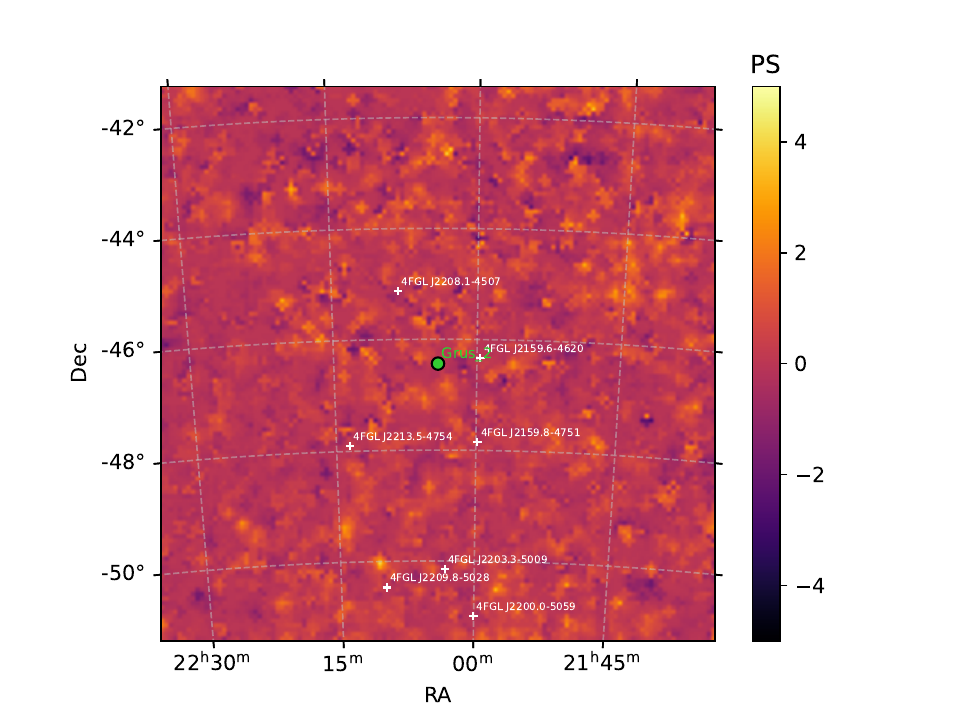}
    \caption{PS Map for the ROI centered on \textit{Grus II} modeled following the prescriptions from \cite{AMD+24} (\textbf{left}), or the ones from this work (\textbf{right}), see \S~\ref{sec:updates} for details.}
    \label{fig:PSMaps_compared}
\end{figure*}

Fig.~\ref{fig:PSMaps_compared} shows the PS Maps for \texttt{Grus II}, when the ROI is modeled following the procedure outlined in \cite{AMD+24} (left), or the one from this work (right).
In this figure, it is evident how the new modeling methods have suppressed the non-random mismatch between model and data, leaving only non-significant ($<4\sigma$) random fluctuations.
In particular, the extended discrepancies observed in the left panel are not present in the new analysis due to the stricter energy cut, and the point-like high-TS (>16) peaks are included in the model of the ROI as point-like sources.

The SED likelihood profiles obtained from these models are converted to the $M_\chi$ vs $\sv$ space assuming the CosmiXs annihilation framework, from \cite{Arina+24}.
Compared to the PPPC4DMID \citep{Cirelli+11} used in some previous analyses of \textit{Fermi} data for DM studies, this framework uses an updated treatment for the electro-weak corrections.
However, in the energy range considered here, the effects on the annihilation spectra are marginal, with an average change of $\sim 0.5\%$ in the $\tau^{+}\tau^{-}$ and $\sim 1\%$ in the $b\bar{b}$ channel over the considered values of $M_\chi$.

\section{Results}
\label{sec:results}

\begin{figure*}
    \label{fig:ULs}
    \centering
    \includegraphics[width = \textwidth]{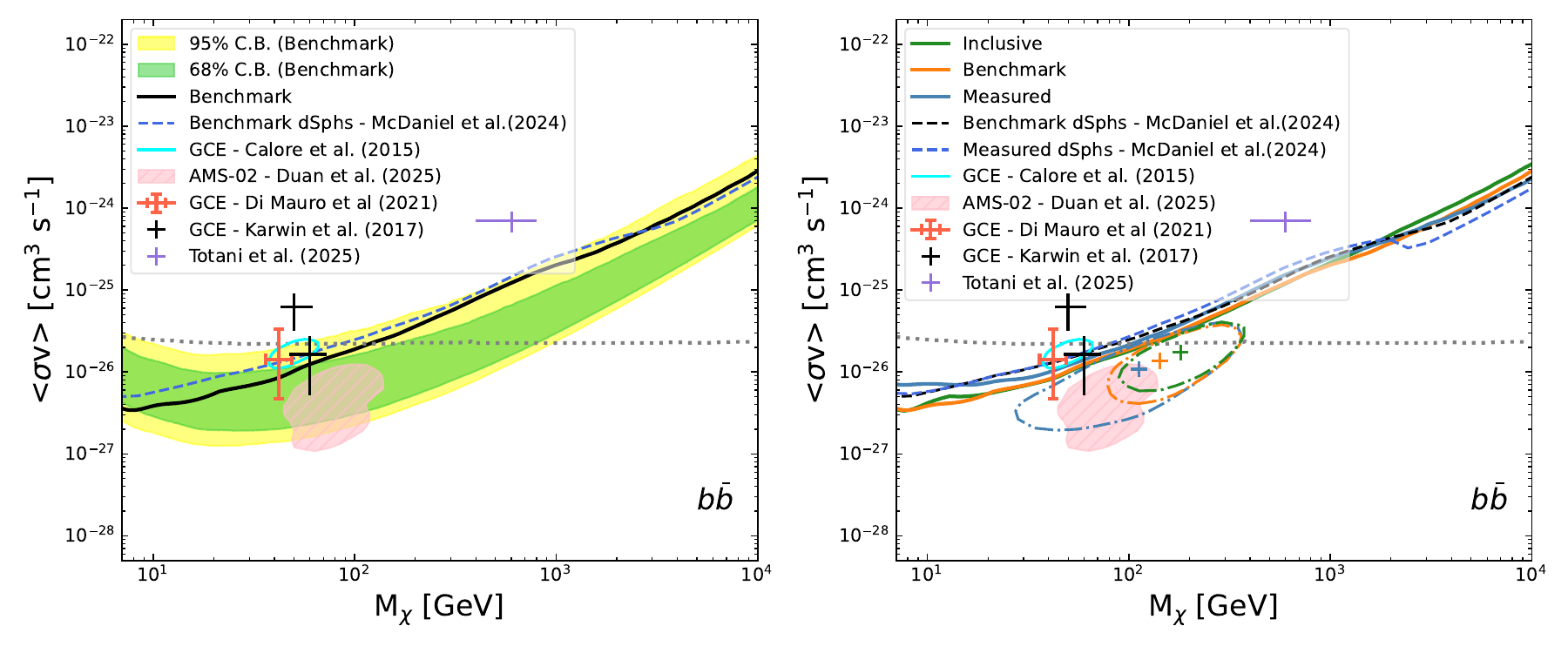}
    \includegraphics[width = \textwidth]{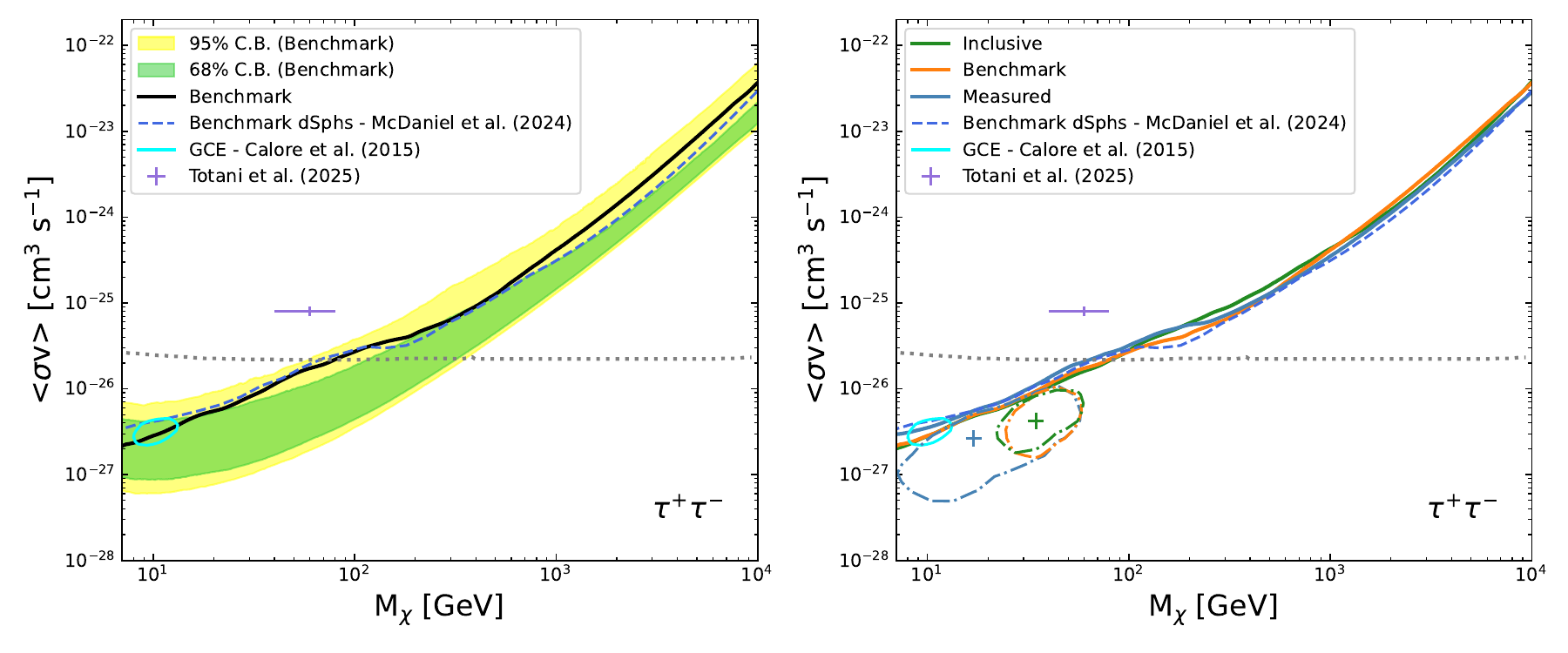}
    \caption{\textbf{Left:} Upper limits for the cross section of annihilation as a function of the mass of the DM particle for the $b\bar{b}$ (\textbf{top}) and $\tau^{+}\tau^{-}$ (\textbf{bottom}) channels, for the benchmark sample. The green and yellow bands are the $68\%$ and $95\%$ containment bands obtained from the blank fields analysis.
    The dashed line represents the upper limits presented in \citet{AMD+24}, while the red point, the black points, and the cyan contour refer to the mass and cross section implied by the Galactic center excess, as reported by \cite{DiMauro+21, Karwin+17} and \cite{Calore+15}, respectively. The purple point refers to the halo-like emission reported by \cite{Totani+25}.
    The pink shaded area represents the values derived from AMS-02 antiproton data. \\
    \textbf{Right}: Upper limits for the cross section of annihilation as a function of the mass of the DM particle in the Measured (blue), Benchmark (orange) and Inclusive (green) dSphs samples, in the $b\bar{b}$ (\textbf{top}) and $\tau^{+}\tau^{-}$ (\textbf{bottom}) channels.
    The dashed lines are the upper limits presented in \citet{AMD+24} for the Benchmark (black) and Measured (blue) samples.
    The points and dot-dashed contours are the coordinates of the point of maximum significance from the combined TS profiles of the three samples of dSphs, and their 1$\sigma$ local confidence level, see \S~\ref{sec:results} for the details.\\
    In all four panels, the dotted grey line is the $\sv$ for thermal relic WIMP DM as a function of $M_\chi$ \citep{Steigman+12}.}
\end{figure*}

\noindent No significant $\gamma$-ray emission is detected at a confidence level of $5\sigma$ or above in any of the dSphs or samples of combined dSphs.

From the combined TS profiles for the three samples of dSphs we estimate the upper limits for $\sv$ as a function of $M_\chi$.
As shown in Fig.~\ref{fig:ULs}, we obtain no significant difference from the results of the upper limits presented in \citet{AMD+24}, indicating that the changes to the analysis method did not affect the sensitivity to DM annihilation signals achieved by this selection of dSphs.
To quantify the significance of the results, for each sample of dSphs, we perform the combined likelihood analysis on 100000 combinations of blank fields, randomly selected without replacement to have the same number of targets as the respective sample.
For each selection, we evaluate the combined TS profile in the $\sv$ vs $M_\chi$ space and select the maximum value labeled as $TS_{max}$ hereafter.
The TS maxima of the combined blank fields profiles deviate from the asymptotical regime of random fluctuations, represented by the $\chi^2$ distribution, due to mismodeling of the background and intrinsic scatter in the properties of each sample.

\begin{figure*}
    \centering
    \includegraphics[width = \textwidth]{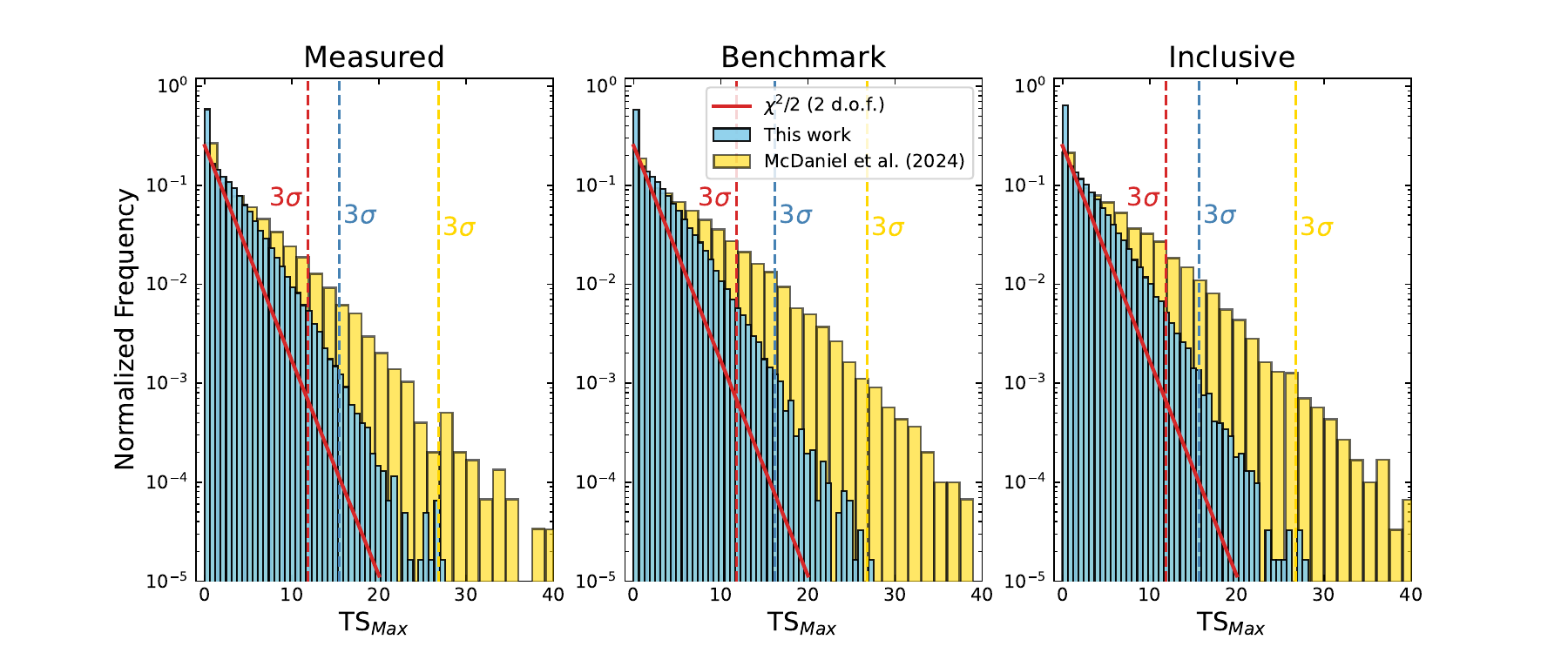}
    \caption{TS distribution for combinations of blank fields in both the $b\bar{b}$ and $\tau^{+}\tau^{-}$ channels of annihilation for the Measured (\textbf{left}), Benchmark (\textbf{centre}) or Inclusive (\textbf{right}) samples.
    The blue histograms refer to the results from this work, while the yellow ones are the ones presented in \citet{AMD+24}. 
    For reference, we report the asymptotic $\frac{1}{2}\chi^2$ (2 d.o.f.) behaviour, following Chernoff's theorem.
    The vertical dashed lines are the Gaussian-equivalent $3\sigma$ confidence levels (99.73\% quantile) for the $\frac{1}{2}\chi^2$ (2 d.o.f.) distribution (red), our work (blue), and \citet{AMD+24} (yellow).}
    \label{fig:bf_ts_histos}
\end{figure*}

\begin{figure*}
    \centering
    \includegraphics[width = \textwidth]{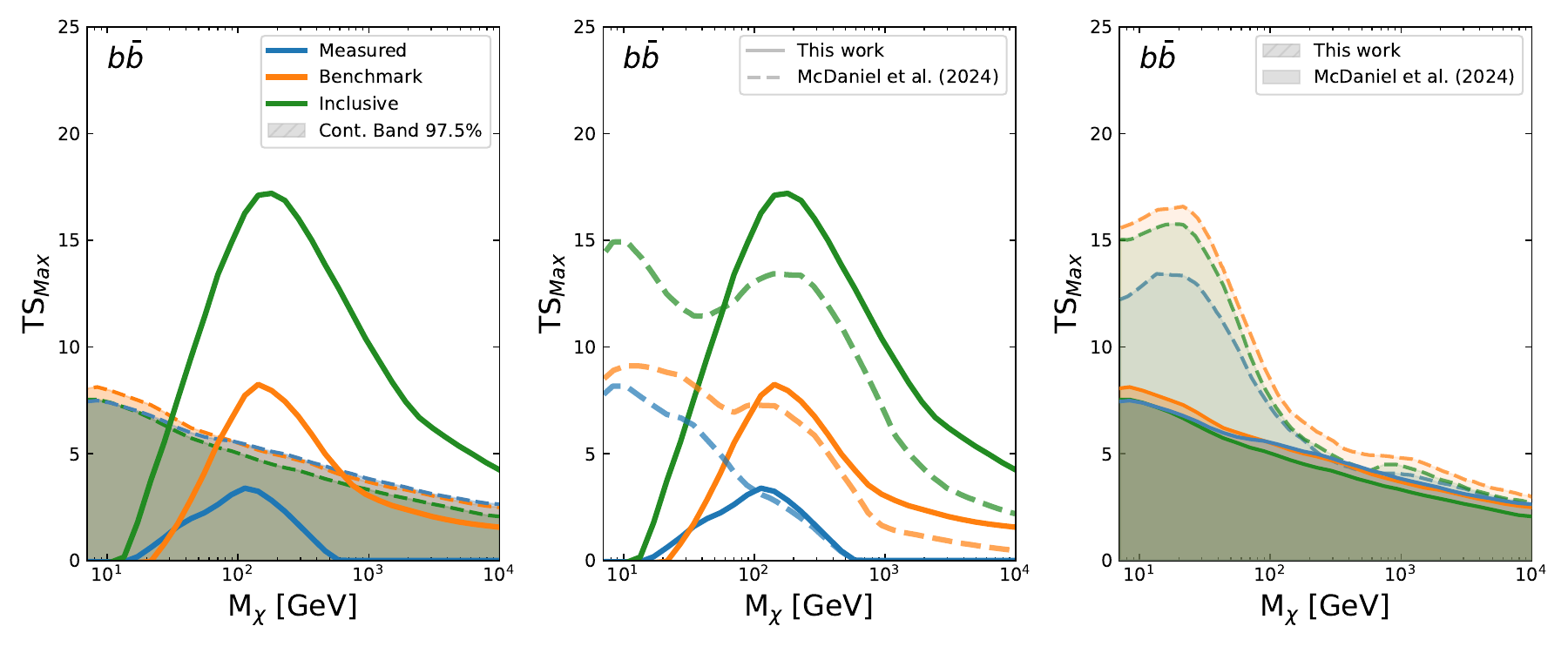}
    \includegraphics[width = \textwidth]{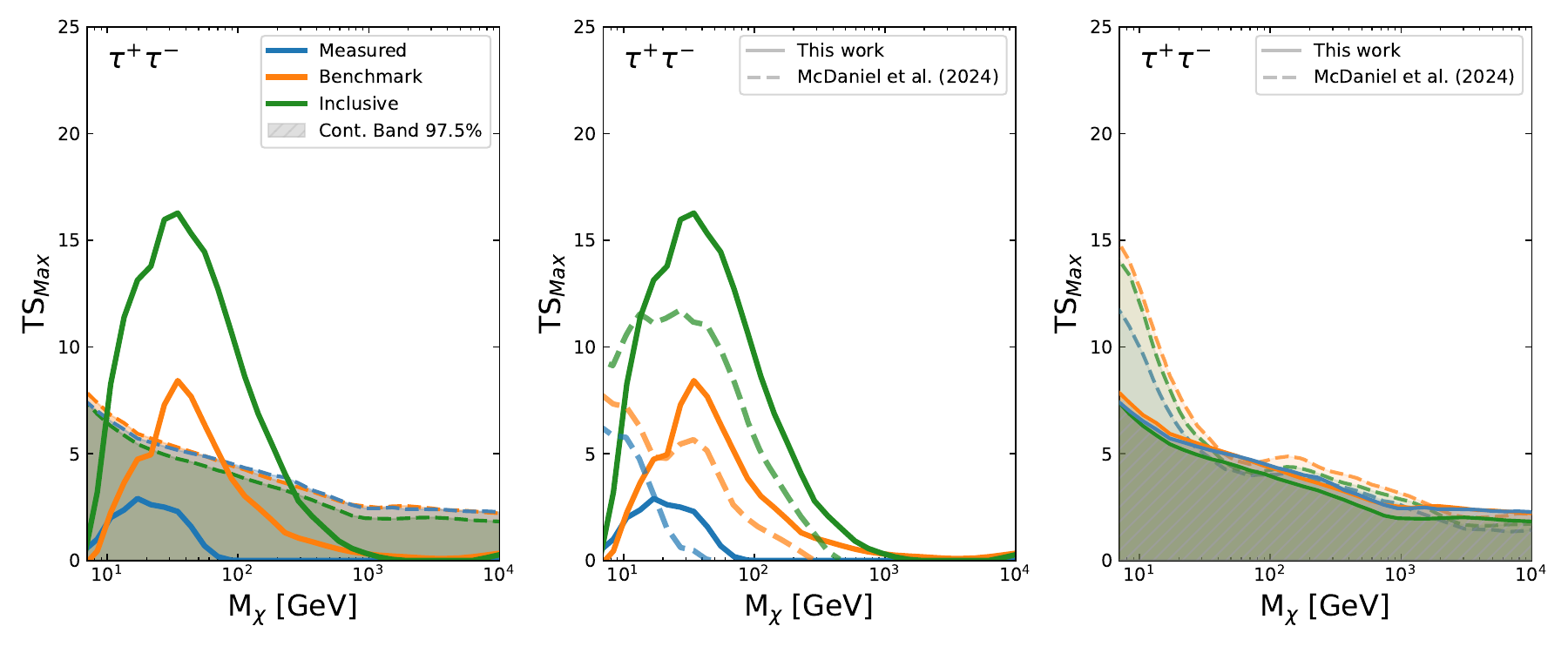}
    \caption{Maximum TS profiles as a function of $M_\chi$ across all cross-section values for the dSphs and $97.5\%$ containment bands from the blank fields analyses, for Measured (blue), Benchmark (orange) and Inclusive (green) samples, in the $b\bar{b}$ (\textbf{top}) and $\tau^{+}\tau^{-}$ (\textbf{bottom}) channels of annihilation.
    The panels on the left compare the dSphs profiles (solid lines) to the containment bands (etched bands) from this work. 
    The central panels compare the dSphs profiles from this work (solid lines) to the ones from \citet{AMD+24} (dashed lines).
    The panels on the right compare the containment bands, obtained from the analysis of the blank fields, from this work (etched bands) to the ones from \citet{AMD+24} (solid bands).}
    \label{fig:ts_vs_mass_combined}
\end{figure*}

Fig.~\ref{fig:bf_ts_histos} compares the distributions of TS maxima of the combined blank fields profiles from this work to the ones presented in \citet{AMD+24} and to the $\chi^2$ distribution, for all three samples.
In all three cases, we obtain a significant improvement, with a strong suppression of the high $TS_{max}$ tails, narrowing the gap to the theoretical limit.
In particular, the TS distributions are now closer to the $ \frac{1}{2}\chi^2$ (2 d.o.f.), which is the asymptotic expectation under Chernoff's theorem \citep{Chernoff+54}.\\
Fig.~\ref{fig:ts_vs_mass_combined} shows the profiles of maximum TS as a function of $M_\chi$ obtained from the combined TS profiles for the three samples of dSphs, in the $b\bar{b}$ and $\tau^{+}\tau^{-}$ channels of annihilation.
In the left panels, the profiles are compared to the $97.5\%$ containment bands obtained from the analyses of the blank fields.
The central panels compare these profiles, obtained from the analysis of the dSphs in this work, to those presented in \citet{AMD+24}, while the panels on the right compare the containment bands, obtained from the analyses of the blank fields.
We see, both in the profiles and the containment bands, a suppression of the low-mass peaks.
This is due partly to the higher energy cut applied to the \textit{Fermi} data, and partly to the adaptive fit of the ROI, which attenuates the mismatches between data and model that are more frequent towards the lower end of the energy interval.
We also notice that the dSphs profiles show no significant changes of the peaks at $\sim 150$ GeV in the $b\bar{b}$ channel (and at $\sim 30$ GeV in the $\tau^{+}\tau^{-}$ channel).
\begin{table*}[ht!]
\centering
\caption{Best values of DM candidate mass and cross section for annihilation, with local and global $p$ values (and significance) for each sample at the local significance peak.}
\label{tab:significance}
\begin{tabular}{lccc}
\hline\hline
\multicolumn{4}{c}{$\bar{b}b$ channel} \\
\hline
Sample & $M_\chi$ [GeV] & $p_{\mathrm{local}}$ & $p_{\mathrm{global}}$ \\
\hline
Measured & 112.5 & $5.8 \times 10^{-2}$ (1.6$\sigma$) & $2.7 \times 10^{-1}$ (0.6$\sigma$) \\
Benchmark & 142.5 & $6.9 \times 10^{-3}$ (2.5$\sigma$) & $5.0 \times 10^{-2}$ (1.6$\sigma$) \\
Inclusive & 180.5 & $9.0 \times 10^{-5}$ (3.7$\sigma$) & $1.6 \times 10^{-3}$ (3.0$\sigma$) \\
\hline\hline
\end{tabular}
\end{table*}
The best values of $M_\chi$ and $\sv$ corresponding to these peaks are reported in Fig.~\ref{fig:ULs} with their 1$\sigma$ uncertainty contours, and in Tab \ref{tab:significance} with their local and global significance.
The global significance is obtained by comparing the TS of the peak with the corresponding blank field distribution from Fig.~\ref{fig:bf_ts_histos}.
These distributions take into account both channels of annihilation being considered, as well as the entire range of possible values for $M_\chi$, therefore correcting the significance for the look-elsewhere effect.
The local significances are obtained from similar distributions, evaluated in one channel of annihilation and at a given $M_\chi$ corresponding to the mass of the peak in TS.

For the Benchmark and Inclusive samples, we observe an increase in both local and global significance in both channels.
In the $b\bar{b}$ channel, the local (global) significance for the Benchmark sample increased from $2.1\;\sigma$ ($0.7\;\sigma$) reported in \cite{AMD+24}, to  $2.5\;\sigma$ ($1.6\;\sigma$) found in this study.
For the Inclusive sample, the previous local (global) significance $3.2\;\sigma$ ($1.6\;\sigma$) from \cite{AMD+24} is here found to be $3.7\;\sigma$ ($3.0\;\sigma$).
In both cases, we observe an increase in the local significance equivalent to the effect of 10 additional years of data, as shown in Fig.~7 of \cite{AMD+24}, and an almost doubled global significance.
Additionally, the peaks are closer together.
\cite{AMD+24} estimates the mass of the DM candidate to be, in the $b\bar{b}$ channel, $\sim180.5$ GeV for the Benchmark sample, and $\sim389.4$ GeV for the Inclusive sample. 
In this study, the expected DM masses show better agreement between the two samples, resulting to be $\sim142.5$ GeV (Benchmark) and $\sim180.5$ GeV (Inclusive).
Similar changes are observed in the $\tau^{+}\tau^{-}$ channel, for both the significances and the predicted masses.
For the Measured sample, the results of \citet{AMD+24} peaked at very low values of $M_\chi \sim 8.4$ GeV in the $b\bar{b}$ channel, and $\sim1.6$ GeV in the $\tau^{+}\tau^{-}$ channel.
With the suppression of the low-mass fluctuations achieved in this study, the global significance is reduced, from $0.8\;\sigma$ to $0.6\;\sigma$ in the $b\bar{b}$ channel, and from $0.6\;\sigma$ to $0.5\;\sigma$ in the $\tau^{+}\tau^{-}$ channel.
However, the peak is shifted closer to the other two samples, with a predicted mass of $M_\chi\sim112.5$ GeV in the $b\bar{b}$ channel, and of $M_\chi\sim17.0$ GeV in the $\tau^{+}\tau^{-}$, maintaining similar local significance, and removing the tension between the three samples, which are now peaking at similar masses in both channels.
Overall, these results (see Tab.~\ref{tab:significance}) indicate that the significance excesses point to $M_\chi$ in the 100-200 GeV range for the $b\bar{b}$ channel and in the 10-40 GeV range for the $\tau^{+}\tau^{-}$ channel.
All three samples are also consistent with the values implied by studies on the GCE, considering the full range of systematics \citep{Calore+15, Karwin+17, DiMauro+21}.

\begin{figure*}
    \centering
    \includegraphics[width = \textwidth]{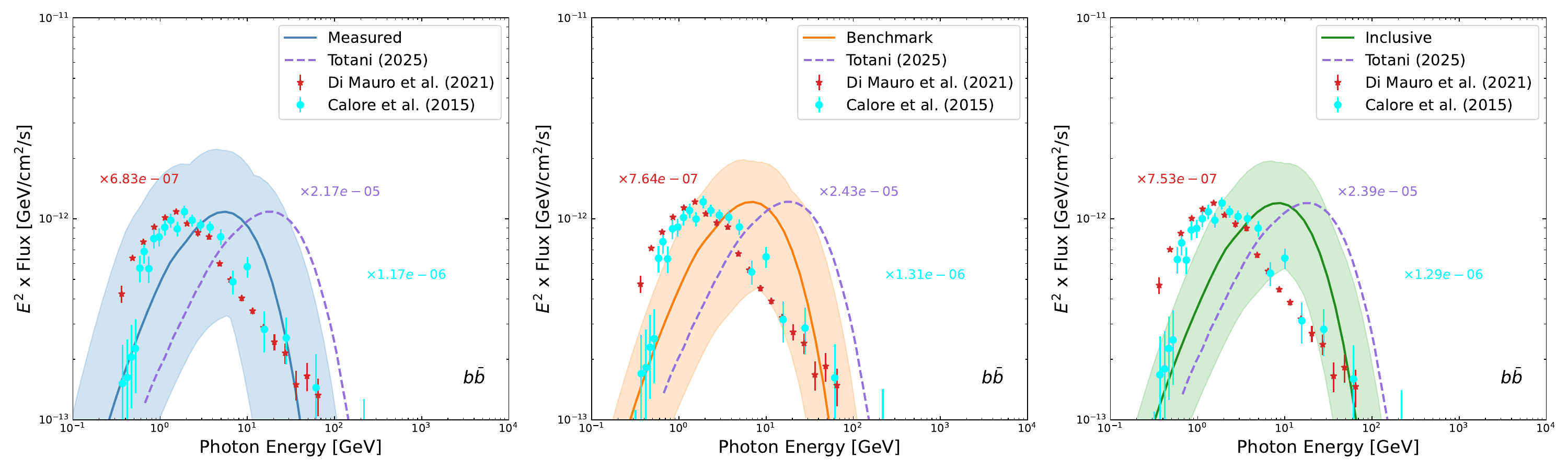}
    \includegraphics[width = \textwidth]{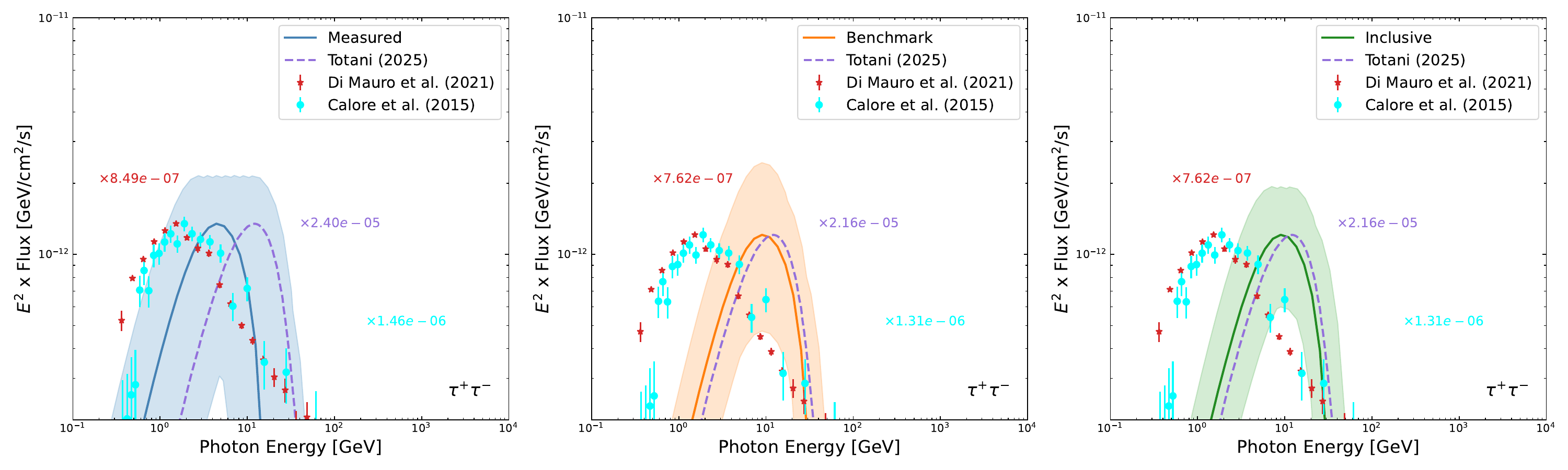}
    \caption{Expected $\gamma$-ray emission from DM annihilation assuming the best-fit $M_\chi$ and $\sv$ for the Measured (\textbf{left}), Benchmark (\textbf{center}), or Inclusive (\textbf{right}) sample, assuming a dSph with the median $J$-factor of each sample, in the $b\bar{b}$ (\textbf{top}) and $\tau^{+}\tau^{-}$ (\textbf{bottom}) channels of annihilation.
    The \textbf{median} $J$-factors are $\log_{10}\bar{J}_M = 18.17$, $\log_{10}\bar{J}_B= 18.23$, and $\log_{10}\bar{J}_I = 18.23$ in the Measured, Benchmark and Inclusive sample, respectively.
    The shaded regions are evaluated from the 1$\sigma$ uncertainties on $M_\chi$ and $\sv$.
    The dashed purple line is the fit of the DM annihilation spectrum to the halo-like excess emission as presented in \cite{Totani+25} (see their Fig. 16, second panel).
    The red and cyan data points are the GCE emissions presented in \cite{DiMauro+21} and \cite{Calore+15}, respectively.
    The line from \cite{Totani+25} and the data from \cite{DiMauro+21} and \cite{Calore+15} are rescaled to match the expected dSphs contribution in each panel at their peak, with the scaling factors reported on the plot in matching color.}
    \label{fig:DM_fluxes}
\end{figure*}

Fig.~\ref{fig:DM_fluxes} shows the expected emission from DM annihilation in dSphs with the median $J$-factor of the Measured ($\log_{10}\bar{J}_M = 18.17$), Benchmark ($\log_{10}\bar{J}_B = 18.23$) or Inclusive ($\log_{10}\bar{J}_I = 18.23$) sample, assuming a DM particle with $M_\chi$ and $\sv$ corresponding to the peak of the combined TS profile of the respective sample (solid line) and their $1 \sigma$ uncertainty (shaded regions).
These curves are compared to the GCE emission interpreted as due to DM presented in \cite{Calore+15} and \cite{DiMauro+21}, rescaled to match the dSphs expected emission at its peak.
The halo-like emission presented in \cite{Totani+25} is also reported, rescaled in the same way.
The scaling factors are reported next to each line, in its respective color.
Within uncertainties, the emissions appear compatible. 
However, we note that the GCE peaks at lower energies and that the $\sv$ obtained by \cite{Totani+25} is a factor of 10 above the upper limits derived from the dSphs.
Yet, the $\sv$ reported by \cite{Totani+25} may be overestimated because it does not take into account the presence of substructures in the MW halo.

\section{Conclusion}
\label{sec:conclusion}
In this work, we explored how the modeling of the background impacts the significance of the results from the analysis of \textit{Fermi}-LAT data coincident with the dSphs.
Through the implementation of stricter cuts on the data and a more careful modeling of the ROIs, we can improve the quality of the model, thereby suppressing the non-random TS fluctuations in the characterization of the null hypothesis, without introducing any bias.
This, in turn, increases the significance of the putative signals found from the combined study of the dSphs.

Achieving a better quality for the models of the background is of particular interest, as the sample of dSphs is expected to double or more over the next decade.
Predictions on the detection capabilities of the Vera C. Rubin Observatory's Legacy Survey of Space and Time (LSST), in fact, show that we could expect $\sim50-100$ additional dSphs to be detected within 10 years \citep{Tsiane+25}.
While \cite{AMD+24} predicts that the excesses in the combined likelihood of the dSphs might break the $5\sigma$ threshold in 10 years with the detection of 65 additional dSphs, the improvements presented in this work will help reach that threshold sooner, or with fewer dSphs.
These excesses imply values of $M_\chi$ and $\sv$ for the DM candidate globally consistent with those derived from the GCE \citep{Calore+15, Karwin+17, DiMauro+21}, although we observe a difference in the spectral shape and most likely energy of the peaks.
Additionally, the predicted emission from DM annihilation in the average member of all three samples of dSphs in this work is compatible, within uncertainties in both spectral shape and energy of the peak, with the halo-like emission presented in \cite{Totani+25}.
Moreover, we find that the values of $M_\chi$ and $\sv$ derived from the antiproton excess observed by the Alpha Magnetic Spectrometer \citep[AMS-02,][]{Duan+2025} are consistent with our results presented in Tab.~\ref{tab:significance}.
However, while the antiproton excess has attracted significant interest due to its compatibility with a thermal relic WIMP of mass $M_\chi \sim 50$--$100$ GeV, its existence remains under debate. In particular, studies such as \cite{Heisig+20} have suggested that the excess may arise from systematic uncertainties in the AMS-02 measurements.
Finally, the recent detections of ultrafaint compact stellar systems \citep{Circiello+25} orbiting the MW, which could be the `darkest' galaxies yet, open up the possibility of some of the newly detected systems having a strong impact on the sensitivity to signals of DM annihilation.

To conclude, the coming years are expected to see crucial developments in the results from searches for $\gamma$-ray signatures of DM annihilation in satellites of the MW, which are among the most promising indirect-search strategies. The analysis of upcoming data will greatly benefit from the methodologies introduced in this work.

\section*{Acknowledgements}
The \textit{Fermi} LAT Collaboration acknowledges generous ongoing support
from a number of agencies and institutes that have supported both the
development and the operation of the LAT as well as scientific data analysis.
These include the National Aeronautics and Space Administration and the
Department of Energy in the United States, the Commissariat \`a l'Energie Atomique
and the Centre National de la Recherche Scientifique / Institut National de Physique
Nucl\'eaire et de Physique des Particules in France, the Agenzia Spaziale Italiana
and the Istituto Nazionale di Fisica Nucleare in Italy, the Ministry of Education,
Culture, Sports, Science and Technology (MEXT), High Energy Accelerator Research
Organization (KEK) and Japan Aerospace Exploration Agency (JAXA) in Japan, and
the K.~A.~Wallenberg Foundation, the Swedish Research Council and the
Swedish National Space Board in Sweden.

MDM acknowledges support from the Research grant TAsP (Theoretical Astroparticle Physics) funded by INFN and from the Italian Ministry of University and Research (MUR), PRIN 2022 ``EXSKALIBUR – Euclid-Cross-SKA: Likelihood Inference Building for Universe’s Research'', Grant No. 20222BBYB9, CUP I53D23000610 0006, and from the European Union -- Next Generation EU.

The work of MASC was supported by the grants PID2024-155874NB-C21 and CEX2020-001007-S, both funded by MCIN/AEI/10.13039/501100011033 and by ``ERDF A way of making Europe''. MASC also acknowledges the MultiDark Network, ref. RED2022-134411-T.
\bibliographystyle{aasjournal}
\bibliography{biblio}

\appendix
\renewcommand\thefigure{\thesection.\arabic{figure}}    
\section{Individual Test Statistics profiles}
\setcounter{figure}{0}    

\begin{figure*}
    \centering
    
    \includegraphics[width = \textwidth]{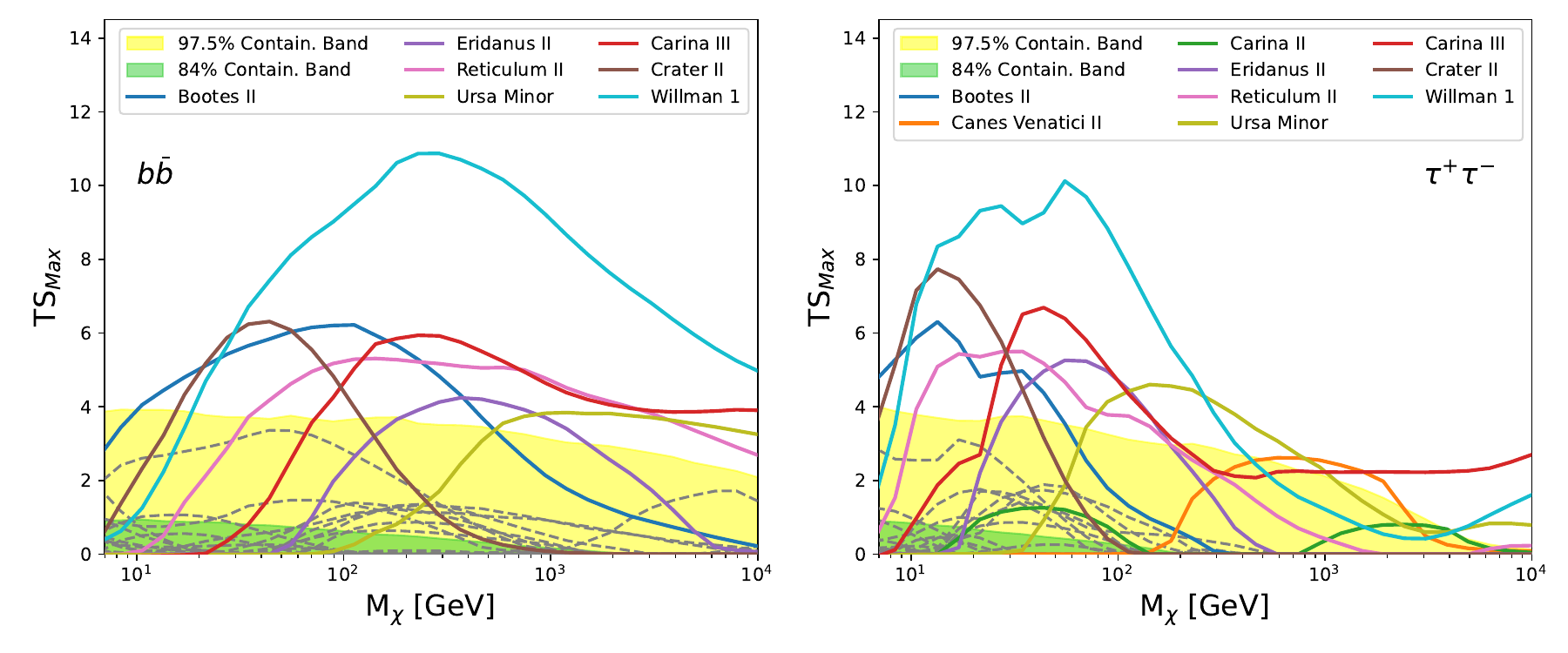}
    \caption{Maximum TS profiles for the individual dSphs as a function of $M_\chi$ across all cross section values, for the $b\bar{b}$ (\textbf{left}) and the $\tau^{+}\tau^{-}$ (\textbf{right}) channels of annihilation.
    The yellow and green bands are the $97.5\%$ and $84\%$ containment bands from the analysis of the individual blank fields.}
    \label{fig:ts_vs_mass_individual}
\end{figure*}

Fig.~\ref{fig:ts_vs_mass_individual} shows the $TS_{max}$ vs $M_\chi$ profiles for the individual dSphs, with the $97.5\%$ containment bands obtained from the analysis of the individual blank fields.
We see again a suppression of the low-mass TS excesses compared to the results presented in \citet{AMD+24}, for both the dSphs profiles and the containment bands.
The profiles for seven dSphs exceed the containment bands in both channels.
Of these dSphs, \texttt{Bo\"otes II}, \texttt{Eridanus II}, \texttt{Reticulum II} and \texttt{Ursa Minor} are bona fide dSphs from the Measured sample, therefore having measured values for the $J$-factor.
\texttt{Carina III} is in the Benchmark sample, with $J$-factor evaluated from its photometric properties through Eq. \ref{eq:photo-J}.
\texttt{Crater II} and \texttt{Willman 1} are from the Inclusive sample.
While both of these dSphs have a measured value for the $J$-factor, \texttt{Crater II} is spatially coincident with a radio source that has no association in the 4FGL-DR4 catalog, and \texttt{Willman 1} shows evidence of non-equilibrium dynamics and tidal stripping \citep{Martin+07}.
Finally, \texttt{Canes Venatici II}, also from the Measured sample, only marginally exceeds the containment bands in the $\tau^{+}\tau^{-}$ channel.

\section{Sensitivity to DM signals}
\label{apdx:sensitivity}
\begin{figure*}[t]
    \centering
    \includegraphics[width = 0.49\textwidth]{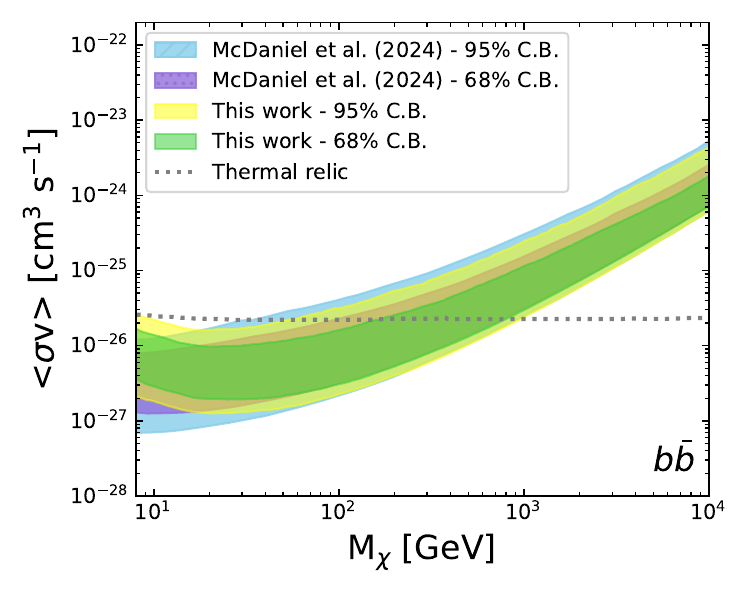}
    \includegraphics[width = 0.49\textwidth]{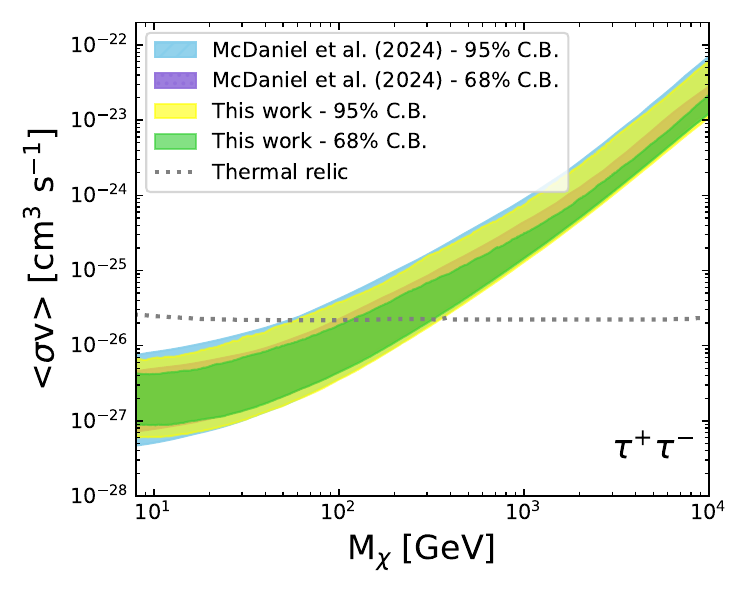}
    \caption{68\% and 95\% containment bands obtained from the blank fields analysis in this work (\textbf{green} and \textbf{yellow}, respectively) compared to the ones obtained in \cite{AMD+24} (\textbf{blue} and \textbf{purple}),  for the $b\bar{b}$ (\textbf{left}) and $\tau^+\tau^-$ (\textbf{right}) channels.}
    \label{fig:cb_comparison}
\end{figure*}

While the choice to restrict the photon selection to higher energies (E > 2 GeV) compared to previous analyses removes most of the model-to-data discrepancies, it could also affect the sensitivity of the sample to DM signals.
To evaluate this effect, in fig.~\ref{fig:cb_comparison} we show the containment bands obtained from the analysis of the blank fields in this work and in \cite{AMD+24}. We find that the new analysis obtains a comparable sensitivity to \cite{AMD+24} across most of the investigated values of $M_\chi$ in both channels. We observe a small deviation at the lowest values of $M_\chi$ in the $b\bar{b}$ channel, where the thermal relic WIMP hypothesis has been largely excluded by previous studies.
Additionally, the containment bands obtained in this analysis are more narrowly distributed around their central value, across all values of $M_\chi$ in both channels.

\newpage
\section{supplementary table}
\setcounter{table}{0}
\renewcommand{\thetable}{\thesection.\arabic{table}}
\begin{longtable}{lcccccccll}
\hline
(1) &(2) &(3) &(4) &(5) &(6) &(7) &(8) &(9) &(10)\\
Name & R.A. (J2000)& Decl. (J2000)& Distance& $r_{1/2}$& $M_V$& $\sigma_{\mathrm{l.o.s.}}$& $\mathrm{log}_{10}J \pm \sigma_J$& Method& Status\\

 & [deg]& [deg]& [kpc]& [pc]& [mag]& [km $\mathrm{s}^{-1}$]& [$\mathrm{log}_{10}\,\mathrm{GeV}^{2}\,\mathrm{cm}^{-5}$]& [M/K/P]& [C/P]\\
\endhead
\hline
\multicolumn{10}{c}{dSphs with Measured $J$-factors}\\
\hline
Aquarius II & 338.48 & -9.33 & 108.0 & 125 & -4.4 & $4.7^a$ & $17.80\;\pm\;0.55^a$& M & C \\ 
Bo\"{o}tes II & 209.51 & 12.86 & 42.0 & 39 & -2.94 & $2.9^a$ & $18.30\;\pm\;0.95^a$& M & C \\
Canes Venatici I & 202.01 & 33.55 & 218.0 & 338 & -8.8 & $7.6$ & $17.42\;\pm\;0.16$& M & C \\ 
Canes Venatici II & 194.29 & 34.32 & 160.0 & 55 & -5.17 & $4.7$ & $17.82\;\pm\;0.47$& M & C \\ 
Carina & 100.41 & -50.96 & 105.0 & 248 & -9.43 & $6.4$ & $17.83\;\pm\;0.10$& M & C \\ 
Carina II & 114.11 & -58.0 & 36.0 & 77 & -4.5 & $3.4$ & $18.25\;\pm\;0.55$& M & C \\ 
Coma Berenices & 186.75 & 23.91 & 44.0 & 57 & -4.38 & $4.7$ & $19.00\;\pm\;0.35$& M & C \\ 
Draco & 260.07 & 57.92 & 76.0 & 180 & -8.71 & $9.1$ & $18.83\;\pm\;0.12$& M & C \\ 
Draco II & 238.17 & 64.58 & 22.0 & 17 & -0.8 & $3.4$ & $18.93\;\pm\;1.54$& M & P \\
Eridanus II & 56.09 & -43.53 & 380.0 & 158 & -7.21 & $7.1$ & $16.60\;\pm\;0.90$& M & C \\ 
Fornax & 39.96 & -34.5 & 147.0 & 707 & -13.46 & $10.6$ & $18.09\;\pm\;0.10$& M & C \\
Grus I & 344.18 & -50.18 & 120.0 & 21 & -3.47 & $4.5$ & $16.50\;\pm\;0.80$& M & P \\
Hercules & 247.77 & 12.79 & 132.0 & 120 & -5.83 & $3.9$ & $17.37\;\pm\;0.53$& M & C \\ 
Horologium I & 43.88 & -54.12 & 79.0 & 31 & -3.55 & $5.9$ & $19.00\;\pm\;0.81$& M & C \\  
Hydrus I & 37.39 & -79.31 & 28.0 & 53 & -4.71 & $2.7^b$ & $18.33\;\pm\;0.36^b$& M & C \\
Leo I & 152.11 & 12.31 & 254.0 & 226 & -11.78 & $9.0$ & $17.64\;\pm\;0.13$& M & C \\ 
Leo II & 168.36 & 22.15 & 233.0 & 165 & -9.74 & $7.4$ & $17.76\;\pm\;0.20$& M & C \\ 
Leo IV & 173.24 & -0.55 & 154.0 & 104 & -4.99 & $3.4$ & $16.40\;\pm\;1.08$& M & C \\ 
Leo V & 172.79 & 2.22 & 178.0 & 39 & -4.4 & $4.9$ & $17.65\;\pm\;0.97$& M & C \\
Pegasus III & 336.1 & 5.41 & 215.0 & 42 & -3.4 & $7.9$ & $18.30\;\pm\;0.93$& M & C \\ 
Pisces II & 344.63 & 5.95 & 182.0 & 48 & -4.22 & $4.8$ & $17.30\;\pm\;1.04$& M & C \\ 
Reticulum II & 53.92 & -54.05 & 30.0 & 31 & -3.88 & $3.4$ & $18.90\;\pm\;0.38$& M & C \\ 
Sagittarius II & 298.16 & -22.07 & 69.0 & 32 & -5.2 & $2.7^c$ & $17.35\;\pm\;1.36^d$& M & P \\ 
Segue 1 & 151.75 & 16.08 & 23.0 & 20 & -1.3 & $3.1$ & $19.12\;\pm\;0.53$& M & C \\ 
Sextans & 153.26 & -1.61 & 86.0 & 345 & -8.72 & $7.1$ & $17.73\;\pm\;0.12$& M & C \\
Tucana II & 342.98 & -58.57 & 58.0 & 165 & -3.8 & $7.3$ & $18.97\;\pm\;0.54$& M & C \\
Tucana IV & 0.73 & -60.85 & 48.0 & 128 & -3.5 & $4.3^e$ & $18.40\;\pm\;0.55^e$& M & C \\
Ursa Major I & 158.77 & 51.95 & 97.0 & 151 & -5.12 & $7.3$ & $18.26\;\pm\;0.28$& M & C \\ 
Ursa Major II & 132.87 & 63.13 & 32.0 & 85 & -4.25 & $7.2$ & $19.44\;\pm\;0.40$& M & C \\
Ursa Minor & 227.24 & 67.22 & 76.0 & 272 & -9.03 & $9.3$ & $18.75\;\pm\;0.12$& M & C \\
\hline
\multicolumn{10}{c}{dSphs with Estimated $J$-factors}\\
\hline
Bo\"{o}tes IV & 233.69 & 43.73 & 209.0 & 277 & -4.53 & - & $17.25\;\pm\;0.60$& P & P \\
Carina III & 114.63 & -57.9 & 28.0 & 30 & -2.4 & $5.6^f$ & $19.70\;\pm\;0.60$& K & C \\
Centaurus I & 189.59 & -40.9 & 116.0 & 76 & -5.55 & - & $18.14\;\pm\;0.60$& P & P \\
Cetus II & 19.47 & -17.42 & 30.0 & 17 & 0.0 & - & $19.10\;\pm\;0.60$& P & P \\
Cetus III & 31.33 & -4.27 & 251.0 & 44 & -2.5 & - & $17.30\;\pm\;0.60$& P & P \\
Columba I & 82.86 & -28.01 & 183.0 & 98 & -4.2 & - & $17.60\;\pm\;0.60$& P & P \\
Grus II & 331.02 & -46.44 & 53.0 & 92 & -3.9 & - & $18.40\;\pm\;0.60$& P & P \\
Phoenix II & 355.0 & -54.41 & 83.0 & 21 & -3.3 & - & $18.30\;\pm\;0.60$& P & C \\
Pictor I & 70.95 & -50.92 & 114.0 & 18 & -3.45 & - & $18.00\;\pm\;0.60$& P & P \\
Pictor II & 101.18 & -59.9 & 46.0 & 47 & -3.2 & - & $18.83\;\pm\;0.60$& P & P \\
Reticulum III & 56.36 & -60.45 & 92.0 & 64 & -3.3 & - & $18.20\;\pm\;0.60$& P & P \\
Tucana V & 354.35 & -63.27 & 55.0 & 16 & -1.6 & - & $18.90\;\pm\;0.60$& P & P \\
\hline
\multicolumn{10}{c}{Special Cases}\\
\hline
Antlia II & 143.89 & -36.77 & 132.0 & 2301 & -9.03 & $5.7^g$ & $16.50\;\pm\;0.60$& K & C \\
Bo\"{o}tes I & 210.02 & 14.51 & 66.0 & 160 & -6.02 & $4.9$ & $18.17\;\pm\;0.30$& M & C \\
Bo\"{o}tes III & 209.3 & 26.8 & 47.0 & 289 & -5.75 & - & $18.65\;\pm\;0.60$& P & C \\
Crater II & 177.31 & -18.41 & 117.5 & 1066 & -8.2 & $2.7^h$ & $15.35\;\pm\;0.26^d$& M & C \\
Horologium II & 49.11 & -50.05 & 78.0 & 33 & -2.6 & - & $18.40\;\pm\;0.60$& P & P \\
Sagittarius & 283.83 & -30.55 & 26.7 & 1565 & -13.5 & - & $19.60\;\pm\;0.20^i$& M & C \\
Virgo I & 180.04 & -0.68 & 91.0 & 30 & -0.33 & - & $18.10\;\pm\;0.60$& P & P \\
Willman 1 & 162.34 & 51.05 & 38.0 & 20 & -2.53 & $4.5$ & $19.53\;\pm\;0.50$& M & C \\
\hline
\caption{List of the 50 dSphs from \cite{AMD+24} used in this analysis.
The first section lists the 30 dSphs that have measured $J$-factors and constitute the Measured sample. 
The second section lists an additional 12 dSphs that have only $J$-factor estimates and together with the Measured section consitute the Benchmark sample. 
The third section lists the Special cases.
The Inclusive sample comprises all dSphs listed in the table.
Column descriptions:
(1) source name
(2) right ascension
(3) declination
(4) heliocentric distance
(5) half-light radius
(6) absolute V-band magnitude
(7) line of sight velocity dispersion
(8) $J$-factor and log-uncertainty on the $J$-factor
(9) method used to determine the $J$-factor
(10) classification status as a confirmed (C) vs probable (P) dSph.
The $J$-factor methods are either to adopt the measured $J$-factors from \cite{Pace+19} where available (M = Measured) or estimate them using the kinematic scaling relation (K = Kinematic, Eq. \ref{eq:kine-J}) or photometric scaling relation (P = Photometric, Eq. \ref{eq:photo-J}) derived in \cite{Pace+19}.
For $J$-factors predicted from scaling relations we assume an error of 0.6 dex.
Column references: (2-6) \cite{Drlica-Wagner+20}, (8,9) \cite{Pace+19} unless indicated by a footnote: (a) \cite{Bruce+23}, (b) \cite{Koposov+18}, (c) \cite{Longeard+20}, (d) \cite{Boddy+20}, (e) \cite{Simon+20}, (f) \cite{Li+18}, (g) \cite{Torrealba+19}, (h) \cite{Caldwell+17}, (i) \cite{Evans+23}.}
\end{longtable}

\end{document}